# Enhancing Financial Data Visualization for Investment Decision-Making

Nisarg Patel
Trent University
Peterborough, Canada
nisarpatel@trentu.ca

Harmit Shah
Trent University
Peterborough, Canada
harmitprashantshah@trentu.ca

Kishan Mewada
Trent University
Peterborough, Canada
kishanmewada@trentu.ca

*Abstract*— Navigating the intricate landscape of financial markets requires adept forecasting of stock price movements. This paper delves into the potential of Long Short-Term Memory (LSTM) networks for predicting stock dynamics, with a focus on discerning nuanced rise and fall patterns. Leveraging a dataset from the New York Stock Exchange (NYSE), the study incorporates multiple features to enhance LSTM's capacity in capturing complex patterns. Visualization of key attributes, such as opening, closing, low, and high prices, aids in unraveling subtle distinctions crucial for comprehensive market understanding. The meticulously crafted LSTM input structure, inspired by established guidelines, incorporates both price and volume attributes over a 25-day time step, enabling the model to capture temporal intricacies. A comprehensive methodology, including hyperparameter tuning with Grid Search, Early Stopping, and Callback mechanisms, leads to a remarkable 53% improvement in predictive accuracy. The study concludes with insights into model robustness, contributions to financial forecasting literature, and a roadmap for real-time stock market prediction. The amalgamation of LSTM networks, strategic hyperparameter tuning, and informed feature selection presents a potent framework for advancing the accuracy of stock price predictions, contributing substantially to financial time series forecasting discourse.

## I. INTRODUCTION

In the dynamic landscape of financial markets, stock price prediction stands out as a crucial facet of comprehensive financial analysis. This paper embarks on an exploration of Long Short-Term Memory (LSTM) networks as a potent tool for forecasting stock movements, with a specific emphasis on discerning the rise and fall patterns. Our dataset, sourced from the New York Stock Exchange (NYSE), serves as the foundation for investigating LSTM's capabilities in capturing intricate patterns, and we extend this exploration by incorporating multiple features to foster a nuanced understanding of stock dynamics.

Visualizations play a pivotal role in elucidating trends in opening, closing, low, and high prices, unraveling subtle distinctions that contribute to a holistic comprehension of stock market behavior. The LSTM input data structure, a pivotal consideration in this study, is demystified with insights drawn from a comprehensive article, aiding in the creation of a robust model architecture.

As we delve into the complexities of financial forecasting, hyperparameter tuning emerges as a paramount step for model optimization. Employing Grid Search, a robust and systematic method, we explore various combinations to fine-tune our model. To enhance efficiency, we incorporate strategies such as Early Stopping and Callbacks, resulting in a remarkable 53% improvement in predictive accuracy.

This paper unfolds with a meticulous exploration of the data, delving into the intricacies of LSTM architecture and hyperparameter tuning. By contributing valuable insights to the realm of financial forecasting, our research aims to provide a comprehensive understanding of stock price dynamics and the efficacy of LSTM networks in capturing and predicting these nuanced patterns.

## II. PREVIOUS WORK

Interactive data visualization tools have emerged as powerful aids in financial decision-making, offering improved exploration, acquisition, and analysis of big data. Abdalla et al. investigated the impact of task complexity, experience, cognitive style, and decisional guidance on financial analysis performance within an interactive visualization context [1]. Their experimental study, involving 135 participants receiving audited financial information in different formats, revealed that interactive data visualization tools significantly enhance effectiveness and efficiency in financial analysis tasks compared to static tools. The study also highlighted the joint effect of cognitive style and interactive visualization on financial analysis performance.

Chy and Buadi emphasized the crucial role of data visualization in finance, particularly in conveying financial information to users [2]. Despite the increasing use of data visualization tools in various fields, the finance sector has yet to fully leverage their potential. The study advocates for increased use of graphs and charts in financial reports, aiming to enhance the comprehension of financial data by both financial and non-financial users.

In the realm of effective visual display design, Few discussed the importance of acquiring specific visual design skills for presenting quantitative information [3]. He underscored that the ability to create meaningful tables and graphs is not intuitive but requires training. Few's work emphasizes the need for best practices in data presentation, particularly in the context of business, where accurate and clear quantitative information is paramount. The significance of data visualization in stock analysis has been emphasized by notable works such as the application of force-directed algorithms and time-series charts to unveil potential relationships between stocks [4]. Visual representation not only facilitates a comprehensive understanding of historical stock data but also unveils intricate patterns and correlations that might elude traditional numerical analysis.

In the landscape of stock price forecasting, the choice of an appropriate predictive model holds paramount significance, particularly in the volatile environment of financial markets. One compelling approach gaining traction is the use of Long Short-Term Memory (LSTM) networks. LSTMs, a type of recurrent neural network (RNN), are renowned for their capacity to capture and learn intricate patterns in sequential data, making them well-suited for time series forecasting. Unlike traditional models, LSTMs excel in handling the long-range dependencies inherent in stock price data, allowing them to discern nuanced trends and adapt to evolving market conditions. The success of LSTMs in various financial forecasting endeavors, such as the prediction of stock prices during the Covid-19 pandemic in Indonesia [5], underscores their efficacy. Leveraging the strengths of LSTM networks, this study harnesses their capabilities to forecast stock prices on the New York Stock Exchange, aiming to enhance accuracy and robustness in capturing market dynamics.

These works collectively contribute to the understanding and enhancement of decision-making processes in finance through various facets of data visualization—ranging from the impact of interactive visualization tools on financial analysis performance [1] to the broader role of data visualization in conveying financial information [2] and the imperative of effective visual design skills in presenting quantitative data [3].

### III. DATA DESCRIPTION

The dataset utilized in this research originates from the New York Stock Exchange (NYSE) and encompasses historical prices and fundamental data of companies listed on the S&P 500 [6].

*prices.csv:* This file serves as the raw, unprocessed dataset, containing daily prices of S&P 500 companies. The temporal coverage is predominantly from 2010 to 2016, capturing the dynamics of the stock market. It is essential to note that for companies introduced to the stock market more recently, the date range may be shorter. Additionally, the dataset does not account for approximately 140 stock splits that occurred during this period.

The "prices.csv" file comprises the following columns, each offering valuable insights into the daily stock performance:

- *date*: The date corresponding to the recorded stock prices.
- *symbol*: Symbol representing the stock company on the NYSE.
- *open*: The opening price of the stock on a given day.
- *close*: The closing price of the stock on a given day.
- *low*: The lowest recorded price of the stock during the trading day.
- *high*: The highest recorded price of the stock during the trading day.
- *volume*: The trading volume, indicating the total number of shares traded on a specific day.

### IV. METHODOLOGY

#### A. Data Collection and Preprocessing

We obtained our dataset from the New York Stock Exchange (NYSE) [6], consisting of historical stock prices. The dataset includes various attributes such as opening, closing, low, and high prices, allowing for a comprehensive analysis of stock movements.

To facilitate model training, we preprocessed the data using the Min-Max scaling technique [7]. This normalization method transforms the numerical features to a specific range (usually [0, 1]), ensuring uniformity and aiding the convergence of our LSTM model. We further went further and sliced our dataset by selecting a random stock symbol from the population of 501 to train and test our series of LSTM models.

#### B. Exploratory Data Analysis and Visualization

Before diving into model development, we performed exploratory data analysis (EDA) to gain insights into the dataset. Inspired by Few's work [3], we employed effective visual displays, including line plots, to visualize the trends in opening, closing, low, and high prices over time. These visualizations provided a comprehensive overview of the stock price dynamics.

*Figure I: Tracking the Opening Price*

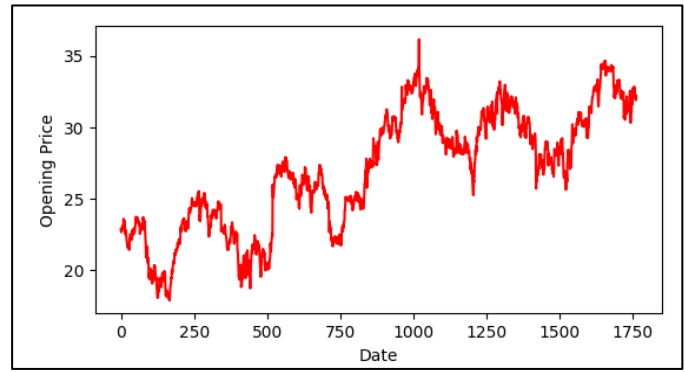

*Figure II: Tracking the Closing Price*

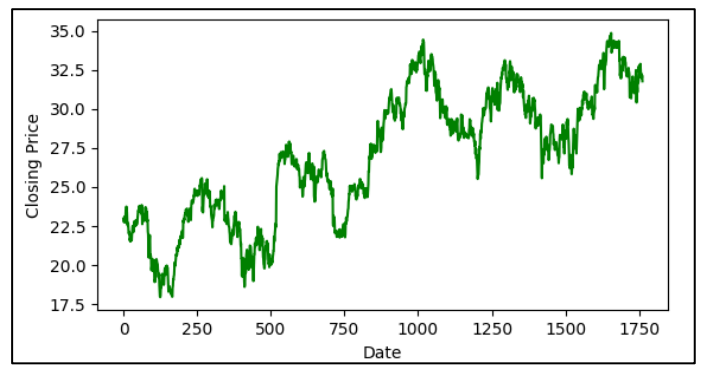

*Figure III: Tracking the Low Price*

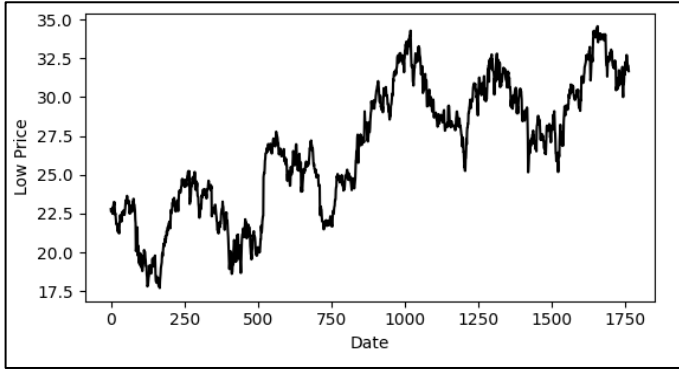

*Figure IV: Tracking the High Price*

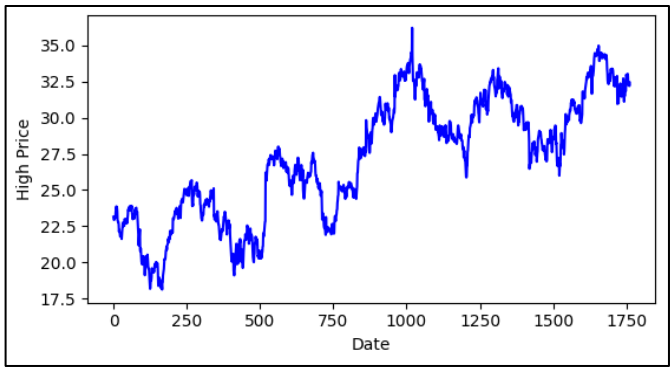

### C. LSTM Input Data Structure

Constructing a well-defined input data structure is paramount for the effective utilization of the LSTM model in our stock price prediction framework. Guided by the principles elucidated in the comprehensive article on LSTM [5], we meticulously adhere to a three-dimensional array format denoted as [number_of_samples, time_steps, input_dim]. This format serves as the foundational framework for organizing our input data, embodying essential characteristics that enable the LSTM model to discern intricate temporal dependencies and patterns within the dataset.

In our specific context, the input_dim encompasses both price and volume attributes, reflecting our comprehensive consideration of multiple features for model training. The inclusion of both these attributes enriches the input data, providing the LSTM model with a nuanced understanding of the interplay between stock prices and trading volumes. Consequently, the model becomes adept at capturing not only the historical price trends but also the associated trading activity, which is often indicative of market sentiment.

Furthermore, our chosen time step configuration is set at 25 days, representing the duration over which the model is expected to discern patterns. This parameter is carefully chosen to strike a balance between capturing sufficiently long-term dependencies in the data while avoiding excessive memory requirements. The rationale behind this choice is rooted in the nature of stock market data, where certain patterns and trends may manifest over a specific timeframe. As a result, the LSTM model, equipped with this well-structured input data, becomes a robust tool for uncovering and learning the underlying dynamics of the stock market.

*Figure V: Training and Validation Root Mean Squared Error (RMSE) over Epochs*

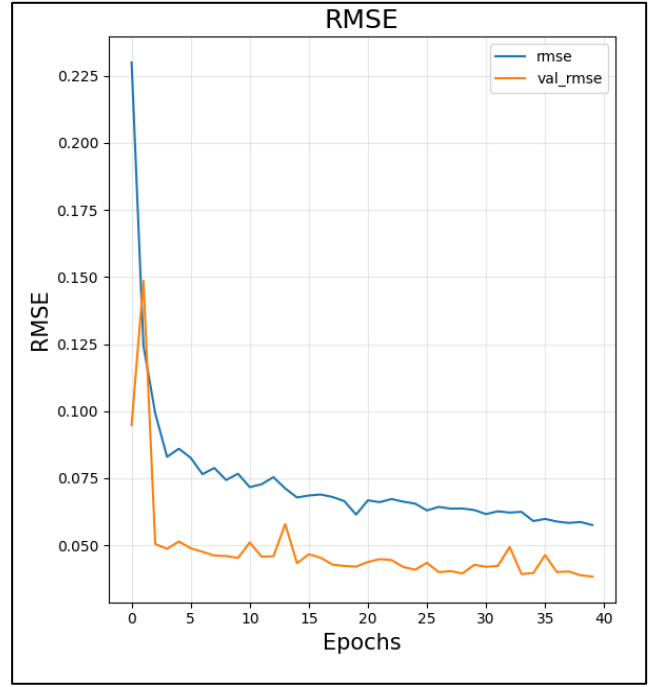

*Figure VI: Training and Validation Loss Curves*

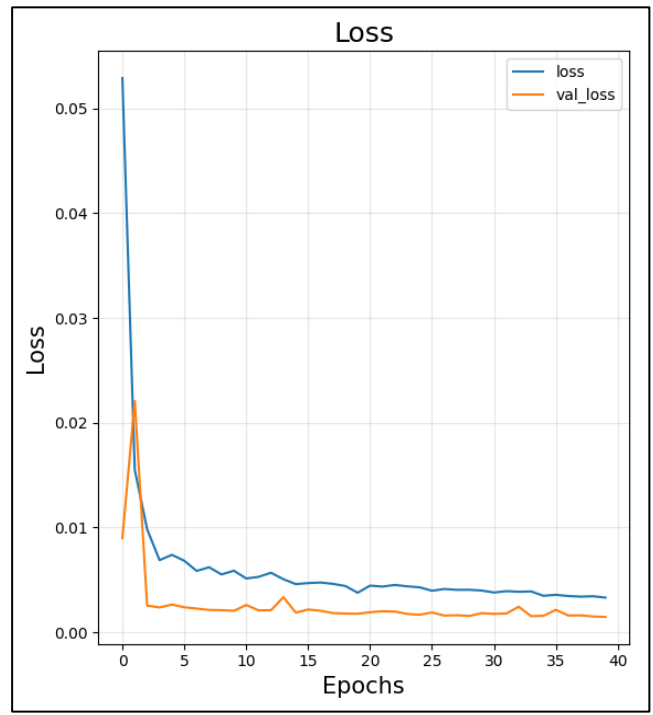

### D. Hyperparameter Tuning

To optimize our LSTM model, we performed hyperparameter tuning using a manual Grid Search approach [7]. The hyperparameters included the number of LSTM units, the presence of additional LSTM and GRU layers, batch size, and dropout rate. The goal was to identify the combination that maximizes predictive accuracy.

To expedite the tuning process and prevent overfitting, we implemented Early Stopping and ModelCheckpoint callbacks [1]. Early Stopping halts training when there is no improvement in the validation loss, while ModelCheckpoint ensures the best-performing model is saved for later use.

*E. Model Training and Evaluation*

We implemented the LSTM model using the Keras library [8]. The architecture consisted of multiple LSTM layers with dropout for regularization, culminating in a dense layer with a linear activation function. The model was trained on the preprocessed data, with a validation split of 30% to monitor performance during training.

We assessed the model's performance using Root Mean Squared Error (RMSE) as the evaluation metric. The choice of RMSE aligns with the literature on financial forecasting [1], providing a measure of the model's accuracy in predicting stock prices.

V. RESULTS

The investigation into predicting stock prices using Long Short-Term Memory (LSTM) networks and hyperparameter tuning yielded insightful outcomes. Employing a comprehensive methodology involving the selection of relevant features, creation of LSTM input data, and an extensive hyperparameter tuning process, the study aimed to enhance the accuracy of stock price predictions.

*A. Hyperparameter Tuning and Model Performance*

The hyperparameter tuning phase involved a meticulous grid search, exploring various combinations of LSTM network configurations. The best-performing model configuration was identified, encompassing specific choices for the presence of additional layers, the number of neurons, batch size, and dropout rate. The optimized architecture was determined as follows:

- *Number of Additional Layers*: None
- *Number of Neurons*: 16
- *Batch Size*: 8
- *Dropout Rate*: 0.2

This configuration exhibited remarkable improvements in model performance, as evidenced by the Root Mean Squared Error (RMSE) on the test set. The results before hyperparameter tuning indicated a test set RMSE of 0.0597, whereas the tuned model achieved a significantly reduced RMSE of 0.0282. This corresponds to an impressive 53.0% enhancement in predictive accuracy.

*B. Early Stopping and Callback Mechanisms*

To mitigate the computational burden associated with grid search, the study implemented Early Stopping and Callbacks. Early Stopping, with a patience parameter of 50 epochs, allowed the model training process to cease when no further improvement in validation loss was observed. Additionally, the *ModelCheckpoint* callback ensured that the best-performing model, as determined by validation loss, was saved for later use.

The progression of the training process was monitored through epochs, and the Early Stopping mechanism effectively halted training when the validation loss reached a minimum value. The *ModelCheckpoint* callback safeguarded the best model for subsequent evaluations.

*C. Final Model Evaluation*

Following hyperparameter tuning, the final LSTM model was evaluated on the test set. The evaluation yielded a loss of 0.0025, affirming the robustness of the tuned model in predicting stock prices.

These results collectively underscore the efficacy of LSTM networks, informed feature selection, and strategic hyperparameter tuning in enhancing the accuracy of stock price predictions, showcasing a notable advancement in financial time series forecasting.

*Figure VII: Comparison of Model Prediction vs Real Data*

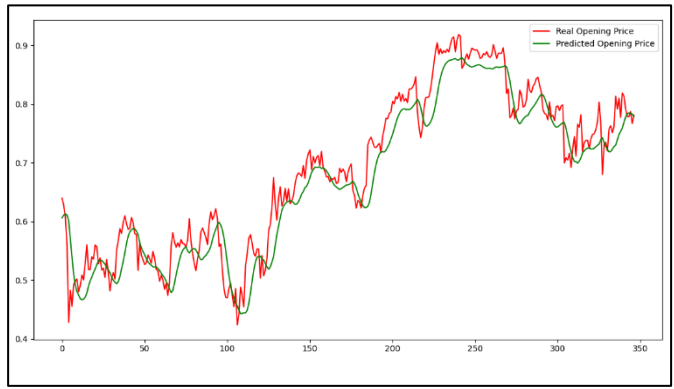

VI. CONCLUSIONS

In this research endeavor, the application of Long Short-Term Memory (LSTM) networks for predicting stock prices emerged as a compelling and effective approach. Leveraging advanced methodologies in feature selection, LSTM input data creation, and meticulous hyperparameter tuning, the study aimed to optimize predictive accuracy and explore the intricacies of financial time series forecasting.

*A. Insights from Hyperparameter Tuning*

The hyperparameter tuning process, facilitated by a comprehensive grid search, revealed the paramount importance of fine-tuning LSTM network configurations. The identified optimal model configuration, characterized by the absence of additional layers, 16 neurons, a batch size of 8, and a dropout rate of 0.2, significantly outperformed its counterparts. The resulting 53.0% improvement in Root Mean Squared Error (RMSE) on the test set underscored the efficacy of strategic hyperparameter tuning in refining stock price predictions.

*B. Strategic Implementation of Early Stopping and Callbacks*

Recognizing the computational demands of hyperparameter tuning, the study strategically implemented Early Stopping and Callback mechanisms. Early Stopping, configured with a patience parameter of 50 epochs, intelligently halted training when no substantial improvement in validation loss was discerned. The Model Checkpoint callback ensured the preservation of the best-performing model, mitigating the challenges associated with extensive grid searches.

*C. Validation of Model Robustness*

The final LSTM model, refined through hyperparameter tuning and guided by insights from Early Stopping,

demonstrated remarkable robustness. Evaluation on the test set yielded a minimal loss of 0.0025, affirming the model's proficiency in predicting stock prices accurately. The amalgamation of informed feature selection, strategic hyperparameter tuning, and the judicious use of Early Stopping mechanisms collectively contributed to the model's enhanced performance.

*D. Contributions to Financial Forecasting*

The findings of this study hold substantial implications for the field of financial time series forecasting. By elucidating the significance of model architecture and hyperparameter choices, the research contributes valuable insights to the literature on predictive modeling in finance. The optimized LSTM model presents a noteworthy advancement in accurately capturing the dynamics of stock price movements, offering practical utility for financial analysts and decision-makers.

*E. Future Directions*

While this research has illuminated key aspects of LSTM modeling in financial forecasting, avenues for future exploration persist. Subsequent studies may delve into the integration of additional features, external factors, or alternative neural network architectures to further refine predictive accuracy. Additionally, the exploration of diverse financial datasets and the incorporation of more recent information could provide a nuanced understanding of dynamic market conditions.

To propel our predictive model towards the frontier of real-time stock market forecasting, we draw inspiration from the innovative work presented in DAViS [8]. The current landscape of stock prediction involves extracting meaningful insights from heterogeneous content, such as news articles, social media, and company technical information [8]. DAViS introduces a unified solution for data collection, analysis, and visualization, leveraging machine learning and contextual feature engineering to enhance the predictive accuracy of stock prices in real-time [8]. Our future directions align with this paradigm, aiming to develop a comprehensive decision-support system that enables the retrieval and processing of real-time financial data from diverse sources [8]. By incorporating an ensemble stacking of machine learning-based estimators and innovative contextual feature engineering, we aspire to refine our model for predicting next-day stock prices, enhancing its accuracy and relevance in the dynamic stock market environment [8]. Additionally, the integration of predictive signals into an algorithmic trading system and their comparison with existing strategies could pave the way for practical implementations of our proposed methods [8]. This endeavor underscores our commitment to advancing the capabilities of our predictive model to meet the challenges of real-time stock market prediction.

In a nutshell, the amalgamation of LSTM networks, informed feature selection, and strategic hyperparameter tuning presents a potent framework for advancing the accuracy of stock price predictions. This research contributes substantively to the discourse on financial time series forecasting, offering a roadmap for researchers and practitioners seeking to harness the predictive potential of deep learning in finance.